\documentclass[12pt,thmsa]{article}

\usepackage{cite}
\usepackage{graphicx}

\begin{document}

\title{Upper and lower bounds to the eigenvalues of an anharmonic oscillator}

\author{Francisco M. Fern\'{a}ndez \thanks{e--mail: fernande@quimica.unlp.edu.ar}\\
INIFTA (UNLP,CCT La Plata--CONICET), Blvd. 113 y 64 S/N, \\
Sucursal 4, Casilla de Correo 16, 1900 La Plata, Argentina}

\maketitle

\begin{abstract}
We obtain tight upper and lower bounds to the eigenvalues of an anharmonic
oscillator with a rational potential. We compare our bounds with results
given by other approaches.
\end{abstract}

\section{Introduction}

Barakat has recently proposed a shifted large--dimension expansion (SLNT)
implemented by means of the Asymptotic iteration method (AIM)\cite{B08}.
According to the author, this alternative approach corrects ``serious
difficulties'' of previous applications of that successful perturbation
theory. He applies this improved SLNT to an anharmonic oscillator with a
rational potential--energy function. The results are not impressive because
the perturbation series are restricted to order six and contain only four
terms becuse the odd coefficients vanish. However, the article title
promises a large--order calculation\cite{B08}.

The purpose of this paper is to calculate some eigenvalues of the anharmonic
oscillator with the rational potential accurately by means of the
Riccati--Pad\'{e} method (RPM) that has proved to be successful for other problems%
\cite{FMT89b,FMT89a,F92,FG93,F95c,F95,F95b,F96,F96b,F97,F08,AF08,F08b}. In Sec. 2 we
show the model and outline the application of the RPM. In Sec. 3 we discuss
the results and compare Barakat's approach\cite{B08} with other applications
of the SLNT.

\section{The Riccati--Pad\'{e} method}

For concreteness we consider the eigenvalue equation
\begin{equation}
-\chi ^{\prime \prime }(x)+\left[ x^{2}+\frac{bx^{2}}{1+cx^{2}}+\frac{l(l+1)%
}{x^{2}}\right] \chi (x)=E\chi (x)  \label{eq:model}
\end{equation}
that applies to one--dimensional and central--field problems. In the former
case we have $l=-1$ and $l=0$ for even and odd states, respectively, and in
the latter case $l=0,1,\ldots $ is the angular--momentum quantum number. One easily
generalizes this equation to take into account a central--field model for
any other space dimension. It is sufficiently general for our present goals. For
comparison purposes we choose $n=0,1,\ldots $ to be the radial quantum number%
\cite{B08}.

The simple function
\begin{equation}
\chi_e(x)=x^{l+1}(1+cx^{2})e^{-x^{2}}  \label{eq:exact_xi}
\end{equation}
is an exact solution to equation (\ref{eq:model}) if
\begin{eqnarray}
b &=&-2c[c(2l+3)+2]  \nonumber \\
e &=&(2l+3)(1-2c)  \label{eq:exact_e}
\end{eqnarray}

The modified logarithmic derivative of the solution
\begin{equation}
f(x)=\frac{l+1}{x}-\frac{\chi ^{\prime }(x)}{\chi (x)}  \label{eq:f(x)}
\end{equation}
can be expanded in a Taylor series about the origin
\begin{equation}
f(x)=x\sum_{j=0}^{\infty }f_{j}x^{2j}  \label{eq:f_series}
\end{equation}
where the coefficients $f_{j}$ are polynomial functions of the energy $E$.

If we require that the rational approximation
\begin{equation}
\lbrack M/N](x^{2})=\frac{\sum_{j=0}^{N+d}a_{j}x^{2j}}{%
\sum_{j=0}^{N}b_{j}x^{2j}}  \label{eq:[M/N]}
\end{equation}
satisfies
\[
\lbrack M/N](x^{2})-\sum_{j=0}^{2N+d+1}f_{j}x^{2j}=\mathcal{O}(x^{2(2N+d+2)})
\]
then the Hankel determinant $H_{D}^{d}$ with elements $f_{d+i+j+1}$, $%
i,j=0,1,\ldots ,N$ vanishes. $D=N+1$ is the determinant dimension and $%
d=0,1,\ldots $. The main assumption of the RPM is that there are sequences
of roots $E_{D}^{d}$, $D=2,3,\ldots $ of the Hankel determinant that
converge towards the eigenvalues of the equation (\ref{eq:model}) for a
given $d$. Previous applications of this approach have shown that the Hankel
series of roots exhibit remarkable rate of convergence and yield accurate
results for a wide variety of problems\cite
{FMT89b,FMT89a,F92,FG93,F95c,F95,F95b,F96,F96b,F97,F08,AF08,F08b}. If $%
E_{D}^{d}<E_{D+1}^{d}$ ($E_{D}^{d}>E_{D+1}^{d}$) then the Hankel series
provide lower (upper) bounds. This property has been proved for some simple
models\cite{FMT89a} but it is more practical to proceed by inspection of the
series.

The RPM yields the exact result in the partially solvable cases; for
example, when $b=-0.46$ and $c=0.1$, then $H_{D}^{d}=(5E-12)^{D-1}P(E)$,
where the roots of the polynomial $P(E)$ give the approximations to the
other eigenvalues. Notice that the RPM yields the exact result $E=12/5=2.4$
for the ground state of the model ($n=l=0$), while, on the other hand, the
AIM with the large--dimension expansion gives this eigenvalue only
approximately\cite{B08}.

\begin{figure}[H]
\begin{center}
\includegraphics[width=9cm]{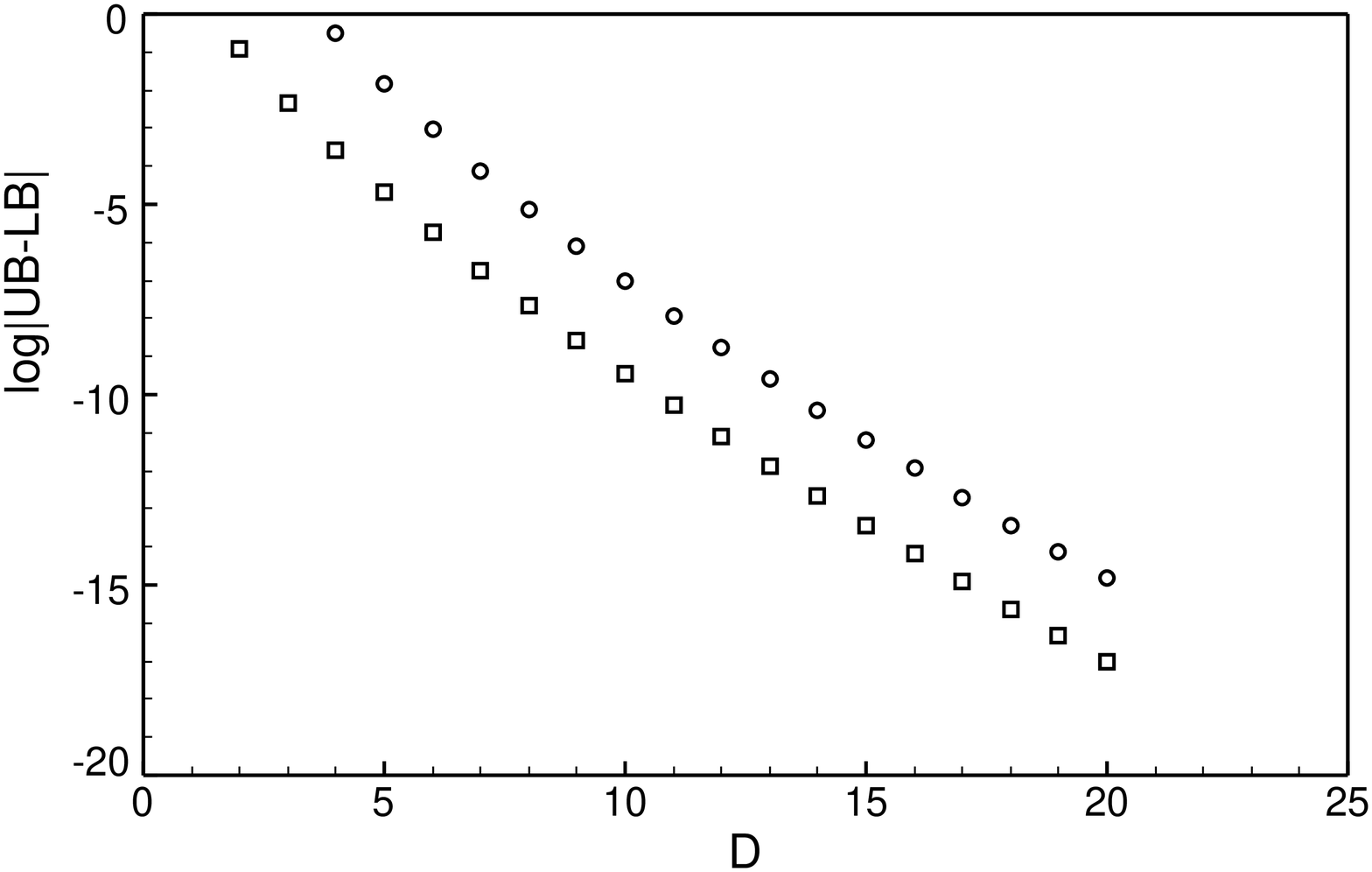}
\end{center}
\caption{Logarithm of the difference between the upper bound and lower bound
for the states with $n=l-1=0$ (squares) and $n=l=1$ (circles) }
\label{fig:Fig1}
\end{figure}

In those cases that the eigenvalue equation (\ref{eq:model}) is not exactly
solvable the RPM provides upper an lower bounds. For concreteness we choose
a model with intermediate values of the parameters: $b=c=1$. Fig. 1 shows $%
\log |E_{D}^{0}-E_{D}^{1}|$, which is is a clear indication of the
convergence of the upper $E_{D}^{0}$ and lower $E_{D}^{1}$ bounds towards
each other, for the states with $n=l-1=0$, and $n=l=1$ . It is well known
that the convergence rate of the Hankel sequences decreases as the number of
radial zeros (given by $n$) increases because we need larger values of $N$
to take into account the increasing number of poles of $f(x)$\cite
{FMT89b,FMT89a,F92,FG93,F95c,F95,F95b,F96,F96b,F97,F08,AF08,F08b}. We clearly
appreciate this fact in Fig. \ref{fig:Fig1}, where the lower curve
corresponds to the state with $n=0$. For this state we obtain the bounds $%
E_{20}^{0}=5.6513933067559477094>E(n=0,l=1)>E_{20}^{1}=5.6513933067559476997$%
. Notice that both the variational $E^{var}=5.65139331725017$\cite
{SHC06} and AIM--SNLT $E^{AIM}=5.654047914$\cite{B08} results lie outside
the bounds and display a misleading number of wrong digits. For the state
$n=l=1$ we conclude that the sixth--order AIM--SNLT
estimate $E^{AIM}=9.708$ lies outside our bounds
$E_{20}^{0}=9.7137541388484906057>E>E_{20}^{1}=9.7137541388484891570$.
We realize that the RPM enables one to estimate the accuracy of other
approaches.

\begin{figure}[H]
\begin{center}
\includegraphics[width=9cm]{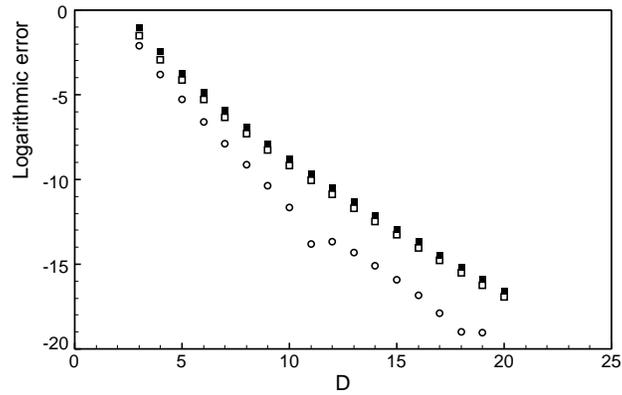}
\end{center}
\caption{Logarithmic error $\log|s_{D+1}-s_D|$ for the sequences $E_D^0$
(squares), $E_D^1$ (filled squares), and interpolation (circles)}
\label{fig:Fig2}
\end{figure}

We can also monitor the convergence rate of a sequence $s_{n},\;n=1,2,%
\ldots $ by means of a sort of logarithmic error $\log |s_{n+1}-s_{n}|$.
Fig. \ref{fig:Fig2} shows this logarithmic error for the sequences $%
E_{D}^{0} $ and $E_{D}^{1}$. We can improve our results by means of an
interpolation of the upper and lower bounds: $i_{D}(p)=\left(
E_{D}^{0}+pE_{D}^{1}\right) /(1+p)$. If $D=K$ is our largest determinant
dimension we obtain the optimal value of $p$ from the condition $%
i_{K}(p)=i_{K-1}(p)$. Fig. \ref{fig:Fig2} shows also the logarithmic error
for the sequence $i_{D}(p)$ and we appreciate what appears to be a
considerably gain of accuracy. From the interpolated sequence we estimate $%
E(n=0,l=1)=5.6513933067559477064$ that is clearly between the bounds given
above.

\section{Conclusions}

We have shown that the RPM yields much more accurate results than the
perturbation version of the AIM. The partial sums of the perturbative AIM
appear to converge towards the RPM results but they may by asymptotic
divergent. It is impossible to state that a series converges by simple
inspection of the first four terms. In the particular case of the anharmonic
oscillator (\ref{eq:model}) the RPM provides tightly accurate upper and
lower bounds to the eigenvalues that are most useful to estimate the
accuracy of the results.

Finally, we want to discuss one of Barakat's comments regarding an early
application of the HPM\cite{MFMC86}. He states ``However, the previous
authors in their work did not give explicit expressions of their algorithm,
each order getting progressively much more complicated than the previous
one, and the derivations were tediously long. Thus, the need arises here to
have a relatively simple, fast and effective method that will provide
large-order shifted $1/N$ expansions.'' This sentence is surprising (to say
the least) because Maluendes et al\cite{MFMC86} gave a recurrence relation
from which one obtains perturbation corrections of any order, either
numerically (in those days) or analytically by the aid of today computer
algebra systems. Despite the fact that each order ``gets progressively more
complicated than the previous one'' Maluendes et al\cite{MFMC86} carried out
perturbation calculations of order twenty five, while Barakat\cite{B08} has
not managed to obtain more that six. Besides, that very same hypervirial
perturbation method has enabled us to derive analytical expressions of the
perturbation coefficients for any order and arbitrary potential--energy
functions\cite{F01}. On the other hand, Barakat presents numerical
perturbation coefficients for some particular states and particular values
of the potential parameters\cite{B08}. I doubt that the AIM may be able to
provide analytical expressions like those just mentioned\cite{F01}. Of
course, the general analytical expressions\cite{F01} are bound to look more
cumbersome than the numbers in Table 1 in Barakat's paper\cite{B08} that
apply to particular cases. It is also important to remember that the AIM
\emph{does not give} results in terms of arbitrary quantum numbers. Any such
result has been obtained by induction from results for $n=0,1,\ldots $\cite
{CHS03,CHS05,CHS05b,BKB06,BB06}. This is possible for extremely simple
problems but it is much more difficult in general. Besides, in those cases
the authors simply verified that their particular results agreed with the
analytical expression derived by other more efficient strategies. Notice
that Barakat presents only the correction of order zero in terms of $n$\cite
{B08}. On the other hand, the unperturbed energy $\varepsilon _{nl}^{(0)}$
appears explicitly in the main equations of the hypervirial perturbative
method and, consequently, all the corrections are straightforwardly given in
terms of it\cite{MFMC86,F01}.

Finally, we mention that a variant of the RPM \cite{AF07} has been
successfully applied to several nonlinear problems of physical
interest\cite{BFG07, BBG08, BBG08b}. It is not clear that the AIM can
be suitable for the treatment of such problems.

\end{document}